\documentclass[graybox, envcountchap]{svmult}

\usepackage{mathptmx}        
\usepackage{amsmath}
\usepackage{amssymb}
\usepackage{color}
\usepackage{helvet}          
\usepackage{courier}         
\usepackage{dirtree}

\usepackage{makeidx}        
\usepackage{graphicx}        
\usepackage{subfig}

\usepackage{multicol}        
\usepackage[bottom]{footmisc}

\usepackage{hyperref}        
\hypersetup{colorlinks=true,urlcolor=blue}

\usepackage[misc]{ifsym}

\makeindex             

\begin{document}


\title{Tip of the Red Giant Branch}
\author{Siyang Li and Rachael L. Beaton}
\institute{Siyang Li (\Letter) \at Department of Physics and Astronomy, Johns Hopkins University, Baltimore, MD 21218, USA, \email{sli185@jh.edu}
\and Rachael L. Beaton \at Space Telescope Science Institute, Baltimore, MD 21218, USA, \email{rbeaton@stsci.edu}}
%
%
\maketitle

\abstract{
While the tip of the red giant branch (TRGB) has been used as a distance indicator since the early 1990's, its application to measure the Hubble Constant as a primary distance indicator occurred only recently. 
The TRGB is also currently at an interesting crossroads as results from the \textit{James Webb Space Telescope (JWST)} are beginning to emerge.
In this chapter, we provide a review of the TRGB as it is used to measure the Hubble constant.
First, we provide an essential review of the physical and observational basis of the TRGB as well as providing a summary for its use for measuring the Hubble Constant. 
More attention is then given is then given to recent, but still pre-JWST, developments, including new calibrations and developments with algorithms. We also address challenges that arise while measuring a TRGB-based Hubble Constant.
We close by looking forward to the exciting prospects from telescopes such as $JWST$ and $Gaia$. 
}


\section{Introduction} \label{sec:introduction}

\subsection{Physical Basis of the TRGB} \label{sec:physics}

The tip of the red giant branch (TRGB) marks the evolutionary transition of stars from the red giant branch (RGB) to the horizontal branch and can be used as a standardizable candle to construct an extragalactic distance ladder to measure $H_0$. After a low mass red giant star ($\sim$ $\leq$ 2 \(M_\odot\)) finishes burning through the hydrogen in its core, it leaves a degenerate Helium core. The star then continues to burn hydrogen in a shell surrounding the Helium core, which increases the core temperature and stellar luminosity. Once the core temperature reaches a critical temperature, helium burning begins. This results in runaway reaction called the Helium Flash \cite{Iben:1983ts}. Here, core degeneracy is lifted, which prevents the star from further increasing in luminosity. After the Helium Flash, the star decreases in luminosity and transitions onto the horizontal branch. The maximum luminosity reached during this evolutionary sequence is a function of initial Helium core mass, which varies very little ($\sim$ 0.001 \(M_\odot\)) for stars from roughly 1.5 to 3 Gyr to $\sim$13 Gyr \cite{Beaton:2018fyo}. This small range of core masses in turn results in a small range of maximum luminosities that allows the TRGB to be used as a standardizable candle. 

\subsection{Observational Basis of the TRGB} \label{sec:observations}

\begin{figure}[b]
\sidecaption
\centering
\includegraphics[scale=.5]{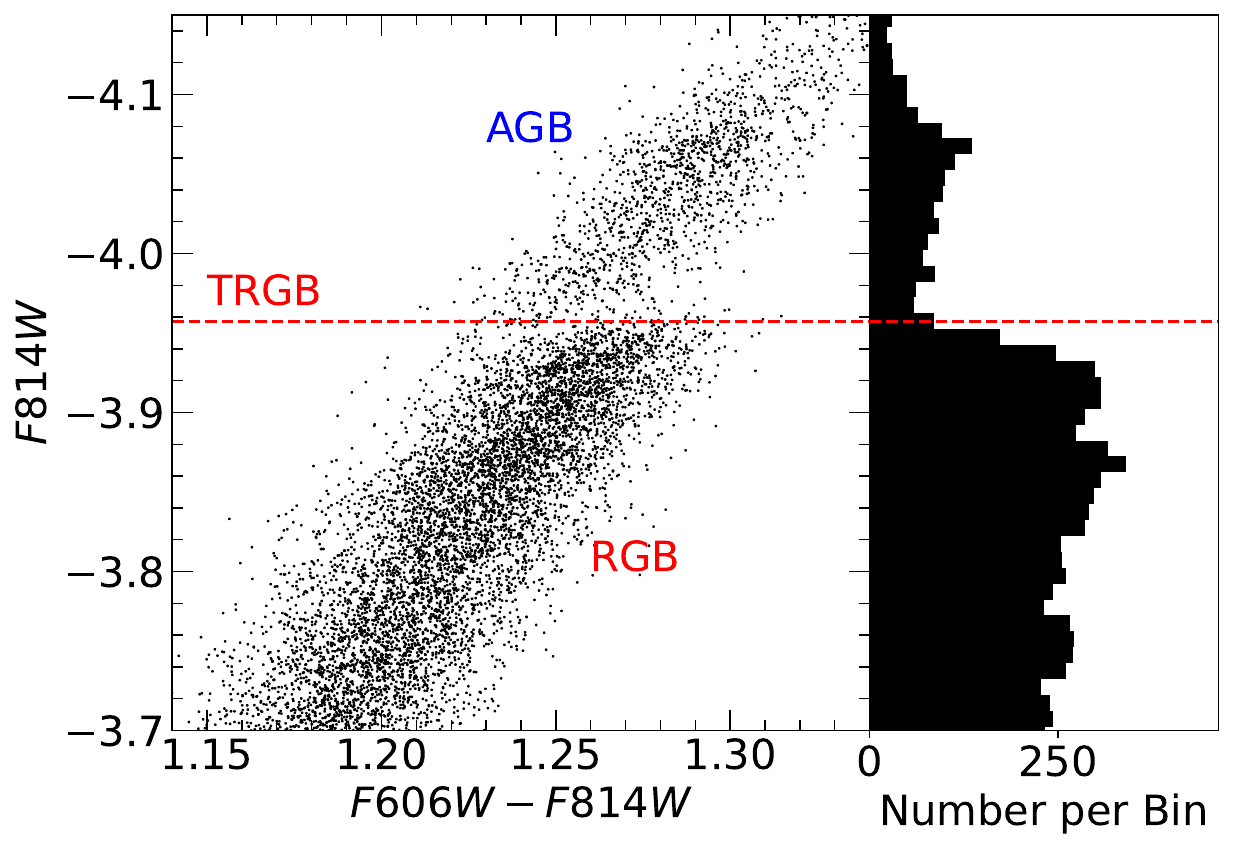}
%
%
\caption{CMD and LF simulated using Artpop \cite{Greco:2022} in the $HST$ $F814W$ and $F606W$ filters and 10 Gyr stars with [Fe/H] = $-1$~dex. We add noise using the noise model from Li et al. 2023 \cite{Li:2023utj}. The location of the TRGB discontinuity is marked with the red dashed line. The RGB and AGB lie below and above the TRGB, respectively.}
\label{fig:TRGB_Simulated_CMD}       
\end{figure}

The increase, then decrease, of a star's luminosity before and after the Helium flash creates a distinct `tip' of the red giant branch that can be visualized in a color magnitude diagram (CMD). To show this, we plot in Fig. \ref{fig:TRGB_Simulated_CMD} a CMD and luminosity function of stars simulated with the Python Artpop package \cite{Greco:2022} and the noise model from Li et al. 2023 \cite{Li:2023utj} in the \textit{Hubble Space Telescope (HST)} $F814W$ and $F606W$ filters. We mark the location of the TRGB with a dashed red line, which corresponds to a discontinuity in the corresponding $F814W$ luminosity function and separates the RGB population below the TRGB from the asymptotic giant branch (AGB) population above the TRGB.  

The TRGB is measured statistically with a population of stars, as the TRGB feature can only be reliably identified and measured when using stars at multiple locations along their evolution leading up to the TRGB. A variety of approaches have been developed to measure the TRGB; we refer the reader to Beaton et al. 2018 \cite{Beaton:2018fyo} for a detailed historical review. Generally, these approaches first apply selection cuts to remove possible contaminants and isolate the red giant branch. These can involve color cuts, which are often based on color calibrations such as from Rizzi et al. 2007 \cite{Rizzi:2007ni} and Jang et al. 2017 \cite{Jang:2017} to isolate metal-poor stars, and spatial cuts involving young star clipping \cite{Wu:2022hxf}, B~mag 25th isophotal radius \cite{Anand:2021sum} or other elliptical cuts \cite{Jang:2021}. We describe algorithms used to measure the TRGB after such cuts in the context of $H_0$ in Sections \ref{subsec:Different Measurement Algorithms} and \ref{subsec:Combining Algorithms and Stellar Populations}. 

Observationally, the TRGB is typically measured in the relatively metal-poor and old halos of galaxies. This is to primarily reduce the effects of contamination from younger stars \cite{Jang:2021} and to avoid internal extinction and crowding. In addition, the TRGB exhibits a metallicity dependence, which can be traced with color due to the line-blanketing effect. The TRGB becomes fainter and brighter at redder colors in the optical and infrared, respectively. Empirically, the TRGB shows the least color dependency in the $I$ or HST $F814W$ filters \cite{Rizzi:2007ni, Jang:2017}; for this reason, these filters are most commonly used to calibrate and measure $H_0$. There is, however, an increasing interest in measuring the TRGB in redder colors to extend TRGB measurements in the Hubble flow and take advantage of the IR capabilities of the \textit{James Webb Space Telescope} ($JWST$). We discuss this further in Section \ref{sec:JWST}.

%



\section{The TRGB Path to $H_0$} \label{sec:Measuring$H_0$}


In this section, we briefly describe how the TRGB can be used to construct a distance ladder to measure $H_0$. We emphasize that this description is not comprehensive and encourage the reader to delve into the literature to learn more about the various methodologies and considerations needed to measure a robust TRGB-based $H_0$.

\subsubsection{Absolute Calibrations} \label{subsec:Absolute Calibrations}


Measuring extragalactic distances with the TRGB to measure $H_0$ requires knowledge of its apparent magnitude, absolute magnitude, and extinction:

\begin{eqnarray}
\label{eq:distance_modulus}
    \mu_0 = m_{host} - M - A
\end{eqnarray}

\noindent where $\mu_0$ is the distance modulus to the host galaxy, $m_{host}$ is the apparent magnitude of the TRGB in the host galaxy, $M$ is the absolute magnitude of the TRGB, and $A$ is the extinction (where we combine foreground and internal extinctions). $m_{host}$ can be directly measured using the methods described later in this chapter. The extinction, $A$, can be obtained from dust maps, for instance from Schlafly \& Finkbeiner 2011 \cite{Schlafly:2010dz}, among others, the discussion of which is outside the scope of this chapter.

With measurements of $m_{host}$ and $A$, we are left with two unknowns, $\mu_0$ and $M$. To obtain $M$, one can invert Equation \ref{eq:distance_modulus} such that 

\begin{eqnarray} \label{eq:distance_modulus2}
    M = m - \mu_0 - A
\end{eqnarray}

\noindent $M$ is intrinsic to the TRGB and varies very little galaxy to galaxy, with variations that can be standardized via measurements of the color and contrast ratio, for instance. This means Equation \ref{eq:distance_modulus2} is general and we can, in principle, exchange $m_{host}$ with a general $m$ and use any galaxy to obtain $M$ provided we also have a direct measurement of its distance, $\mu_0$, that is obtained independently of the TRGB. Examples of galaxies with distances measured geometrically and independently from the TRGB that are most commonly used to calibrate, or `anchor', the TRGB are the Milky Way, Large Magellanic Cloud (LMC), Small Magellanic Cloud (SMC), and NGC 4258. Distances to Milky Way red giants can be determined via parallaxes, the most precise of which have been measured by the $Gaia$ mission \cite{Gaia:2016zol}. The LMC, SMC, and NGC 4258 have geometric distances measured via eclipsing binaries \cite{Pietrzynski:2013gia, Graczyk:2020} and water masers \cite{Reid:2019tiq}, respectively. For derivations of the LMC, SMC, and NGC 4258 distances using these systems, we refer the reader to their respective references. 


With measurements of the geometric distance to and apparent TRGB magnitude in the anchor galaxy (as well as an estimate of the extinction, $A$), one can solve for the absolute magnitude, $M$, of the TRGB using Equation \ref{eq:distance_modulus2}. With $M$ as well as $m_{host}$ and $A$ for the host galaxy, one can then return to Equation \ref{eq:distance_modulus} and solve for $\mu_0$.

We briefly pause to highlight differences between two distinct approaches of calibrating the TRGB and how they affect $H_0$ measurements. The first approach is to minimize uncertainties in the TRGB by applying particular selections. For instance, Anderson et al. 2023 \cite{Anderson:2023aga} utilize the variability of Optical Gravitational Lensing Experiment (OGLE) Small Amplitude Red Giants (OSARGS) \cite{Kiss:2003wk, Wray:2003sm, Soszynski:2004wn, Soszynski:2009wd} in the LMC to achieve a 1.39$\% $ calibration of the TRGB. Another instance can be seen in Hoyt 2021 \cite{Hoyt:2021irv}, who selects 5 out of 20 LMC fields that have more symmetric Sobel responses and narrower bootstrap widths to achieve a 1.5$\%$ calibration. While both these methods are effective at lowering uncertainties in the TRGB calibration to the 1$\%$ level, it is important to note that these calibrations are distinct from calibrations that can be used (at the moment) to measure $H_0$, as these selections are not currently feasible in SN Ia host galaxies with the current resolution of space telescopes such as $HST$ and $JWST$. For instance, the amplitudes of red giants are the same order magnitude of the photometric uncertainties in many SN Ia host galaxies, which makes it difficult to differentiate intrinsic and extrinsic variability. Li et al. 2023 demonstrate in their Appendix D that the same bootstrap width criterion of 0.025~mag used to reject 15 out of 20 LMC fields used in Hoyt 2021 \cite{Hoyt:2021irv} would eliminate all TRGBs measured in SN Ia host galaxies, precluding its use to measure $H_0$ \cite{Li:2023utj}. The key concept here is that similar selections should be applied to both the anchor and host galaxies. Inconsistent selections can introduce biases, such as mismatch of the calibration due to different stellar populations.

The second approach applies the same or similar measurement and selection criterion to both the anchor and host galaxies, even if this results in a higher uncertainty in the calibration.  Here, the calibration and selection of the TRGB remain consistent across the distance ladder such that the calibration more closely reflects the measured luminosity of the TRGB in the host galaxy such as in Scolnic et al. 2023 \cite{Scolnic:2023mrv}. We emphasize that both these approaches are very useful for advancing understanding of the TRGB. One should be careful, however, to be aware of these two conceptually different approaches when choosing a calibration to anchor distance and $H_0$ measurements. We look forward to future instrumentation that can provide the resolution that erases this distinction.

\subsubsection{Calibrating SN~Ia}

Now with a calibration of the TRGB, one can measure distances to galaxies using Equation \ref{eq:distance_modulus}. To measure $H_0$, one in particular would want to measure TRGB distances to galaxies that contain both a well-observed TRGB as well as a luminous standard candle that can reach further into the Hubble flow, such as SN~Ia, to obtain a calibration of the more luminous standard candle. The Tully-Fisher relation is another way one can reach far into the  Hubble flow; we refer the reader to the corresponding section in this book for more information. For simplicity, we will refer to SN~Ia when describing the distance ladder for the remainder of this chapter.

A TRGB distance measurement to a galaxy, combined with a measurement of the apparent magnitude of SN~Ia (such as from Pantheon$+$ \cite{Scolnic:2021amr} or the Carnegie Supernovae Project  \cite{Hamuy:2005tf, Krisciunas:2017yoe}), allows one to calibrate the SN~Ia luminosity using the same principles described in Section \ref{subsec:Absolute Calibrations}. A luminosity calibration of SN~Ia can then be combined with apparent magnitude SN~Ia measurements to measure distances to galaxies further in the Hubble flow where the TRGB is too faint to be observed.

One illustrative example of how to calibrate SN~Ia is described in Scolnic et al. 2023 \cite{Scolnic:2023mrv}. They first find the difference in magnitude between the apparent TRGB and mean SN~Ia apparent magnitude in a given galaxy:

\begin{eqnarray}
\label{eq:SNIa_Cal_1}
    \Delta S = m_{I, TRGB} - m_B^0
\end{eqnarray}

\noindent where $\Delta S$ is the difference between the apparent TRGB and SN~Ia magnitudes, $m_{I, TRGB}$ is the apparent TRGB magnitude, and $m_b^0$ is the apparent SN~Ia magnitude. They then obtain the difference between the absolute magnitudes of the TRGB and SN~Ia by taking the mean $\Delta S$ from all SN~Ia host galaxies and subtracting it from the TRGB luminosity:

\begin{eqnarray}
\label{eq:SNIa_Cal_2}
    M_B^0 = M_{I, TRGB} - \Delta \bar S
\end{eqnarray}

\noindent to obtain the luminosity of SN~Ia, $M_B^0$. We note that Equation \ref{eq:SNIa_Cal_2} is essentially an algebraic rearrangement of Equation \ref{eq:distance_modulus} for two standard candles to the same galaxy. $\mu_0 = m - M$ with TRGB should yield the same distance when using $\mu_0 = m - M$ with SN~Ia when both are in the same galaxy. By plugging in $\Delta S$ from Equation \ref{eq:SNIa_Cal_1} into Equation \ref{eq:SNIa_Cal_2}, one can rearrange to find $(m - M)_{TRGB} = (m - M)_{SN~Ia}$.

\subsubsection{Measuring $H_0$}

Measuring $H_0$ requires knowledge of both redshifts and distances to galaxies. The derivations of the equations used to measure $H_0$ are outside the scope of this chapter; we instead briefly mention that one can conveniently obtain $H_0$ using the Hubble diagram intercepts from previous studies. With a calibration of SN~Ia obtained with the method from the previous section, one can use the equation:

\begin{eqnarray}
\label{eq:logH0}
    log H_0 = 0.2 M^0_B + a_B + 5
\end{eqnarray}

\noindent where $M_B^0$ is the SN~Ia luminosity, and $a_B$ is the intercept of the Hubble diagram at low redshift and can be found from various studies, such as Riess et al. 2022 \cite{Riess:2021jrx}. Equation \ref{eq:logH0} can also be used to measure $H_0$ without $a_B$ by comparing distances relative to another $H_0$ measurement.

\subsection{Measurements of $H_0$ with the TRGB} 

In this section, we highlight a few $H_0$ measurements from the past couple of years. We note that this discussion is not exhaustive and encourage the reader to read the references within these studies for more examples of TRGB-based $H_0$ measurements. 

\begin{enumerate}

\item The most recent $H_0$ measurement from the Carnegie-Chicago Hubble Program (CCHP) results in $H_0 = 69.8 \pm 0.6$ (stat) $\pm$ 1.6 (sys) km s$^{-1}$ Mpc$^{-1}$ \cite{Freedman:2021ahq}. This work updates the Freedman et al. 2019 \cite{Freedman:2019jwv} measurement, which used a Monte Carlo Markov Chain approach and 18 SN~Ia calibrators to measure $H_0$. In this update, Freedman 2021 \cite{Freedman:2021ahq} replaces the Freedman et al. 2019 \cite{Freedman:2019jwv} LMC only calibration with a combination of calibrations from NGC 4258 \cite{Jang:2021}, Milky Way globular clusters \cite{Cerny:2020inj}, LMC \cite{Hoyt:2021irv}, and SMC \cite{Hoyt:2021irv}. They use a Sobel-filter based TRGB measurement methodology which will be described in Section \ref{subsec:Different Measurement Algorithms}.

\item Anand et al. 2022 \cite{Anand:2021sum} independently reduced the data used by CCHP to find $H_0 = 71.5 \pm$ 1.8 km s$^{-1}$ Mpc$^{-1}$. They incorporate $HST$ NGC 4258 observations that were not used in Freedman 2021 to anchor their distance ladder and notably found that they could not reliable identify the TRGB in 4 out of the 15 hosts used by Freedman et al. 2019.  Anand et al. 2022 \cite{Anand:2021sum} also used a model based least-squares fit instead of a Sobel method, a different photometry pipeline, and apply color corrections from Rizzi et al. 2007 \cite{Rizzi:2007ni}. They find a $-0.028$~mag difference (Extragalactic Distance Database - CCHP) distance scales and did not find a significant difference in their $H_0$ measurement when using SN~Ia from the Carnegie Supernovae Project \cite{Hamuy:2005tf, Krisciunas:2017yoe} or Pantheon \cite{Pan-STARRS1:2017jku}.

\item Scolnic et al. 2023 \cite{Scolnic:2021amr} recently measured $H_0 = 73.22 \pm 2.06$ km s$^{-1}$ Mpc$^{-1}$. In their measurement, Scolnic et al. 2023 \cite{Scolnic:2021amr} correct their measured TRGBs to a fiducial contrast ratio to improve standardization across rungs in the distance ladder. They calibrate their TRGB to NGC 4258 \cite{Li:2023utj} and include three new SN~Ia in their analysis that were not previously used in Freedman 2021 or Anand et al. 2022  \cite{Freedman:2021ahq, Anand:2021sum}, in addition to applying peculiar flow corrections and the Pantheon$+$ sample. They find that applying contrast ratio corrections to their TRGB measurements increases the Freedman 2021 \cite{Freedman:2021ahq} and Anand et al. 2022 \cite{Anand:2021sum} $H_0$ by 1.4 and $-$0.3 km s$^{-1}$ Mpc$^{-1}$, respectively. The largest differences, however, came from different treatments of SN~Ia, which would change the Freeman 2021 \cite{Freedman:2021ahq} and Anand et al. 2022 \cite{Anand:2021sum} $H_0$ by 2 and 1.3 km s$^{-1}$ Mpc$^{-1}$, respectively. Their algorithm and data are publicly available via the link provided in Section \ref{sec:Open_Datasets}.

\item Dhawan et al. 2023 \cite{Dhawan:2022gac} used a hierarchical Bayesian SED model, BayeSN, to infer SN~Ia distances calibrated with Cepheids and TRGB to find $H_0 = 70.92$ $\pm$ 1.14 (stat) $\pm$ 1.49 (sys) km s$^{-1}$ Mpc$^{-1}$ with the TRGB calibration. Their hierarchical Bayesian SED model allowed them to simultaneously model optical and NIR SN Ia light curves, decreasing $H_0$ uncertainty by $\sim15\%$ compared to their optical only case.

\item Dhawan et al. 2022 \cite{Dhawan:2022yws} measured $H_0 = 76.94 \pm 6.4$ km s$^{-1}$ Mpc$^{-1}$ using only SN~Ia from the from the Zwicky Transient Facility \cite{Graham:2019qsw} to minimize systematic from host-galaxy bias and different treatments of photometry. They use a single host galaxy NGC 7814 containing SN~Ia ZTF SN Ia SN 2021rhu and measure the TRGB with the same pipeline from CCHP. The large uncertainty is primarily due to their single SN~Ia calibrator; they note that future Zwicky Transient Facility observations of SN~Ia combined with $JWST$ observations of TRGB will significantly improve this measurement.

\item  Blakeslee et al. 2021 \cite{Blakeslee:2021rqi} tied Cepheid and TRGB calibrations to surface brightness fluctuation measurements of 64 galaxies ranging from 19 to 99 Mpc to find $H_0 = 73.3 \pm 0.7 \pm 2.4$ km s$^{-1}$ Mpc$^{-1}$. They obtain their sample from a variety of programs using the $HST$ WFC3/IR $F110W$ filter and four different treatments of galaxy velocities. 

\end{enumerate}
\section{Challenges for $H_0$ Measurements} \label{sec:2}

\subsection{Different Measurement Algorithms} \label{subsec:Different Measurement Algorithms}

The TRGB can be measured using a variety of different methodologies, each of which can introduce challenges for ensuring consistency along the distance ladder to measure $H_0$. Here, we focus on describing the Sobel filter and least-squares fit methods, the two of which are most commonly used to measure TRGB $H_0$.


\subsubsection{Sobel Filter}

The Sobel filter was first used to measure the TRGB by Lee et al. 1993 \cite{Lee:1993jb} and has since been adopted in several different forms. For a detailed review of these variations, we refer the reader to Beaton et al. 2018 \cite{Beaton:2018fyo}. Here, we focus on describing the primary method adopted by the Carnegie-Chicago Hubble Program (CCHP) \cite{Hatt:2017rxl}, which has been used extensively to measure $H_0$. An unsupervised variant of this method has been developed by the Comparative Analysis of TRGBs (CATs) team \cite{Wu:2022hxf, Scolnic:2023mrv, Li:2023utj}; we describe this method separately in Section \ref{subsec:Combining Algorithms and Stellar Populations}.

The Sobel filter concept originates from the field of computer vision and is used to identify edges by evaluating the first derivative of intensities. In computer vision, this may involve finding the edges of an object in a 2D image to isolate the object from its surroundings. For a 1D luminosity function around the TRGB, the Sobel filter can  be used to locate the TRGB discontinuity in the luminosity function with the maximum first derivative. To use this method, one first finely bins the luminosity function around the TRGB in bin widths of order 0.005~mag. Then, the binned luminosity function is smoothed using Gaussian-windowed, Locally Weighted Scatterplot Smoothing (GLOESS) \cite{Persson:2004, Monson:2017,Hatt:2017rxl} to suppress false detections caused by noise. Poisson weights are also applied to de-weight sparsely populated bins. Finally, a Sobel kernel of the form [-1, 0, 1] is evaluated across the luminosity function to compute the first derivatives of the smoothed and weighted luminosity function. The TRGB is taken to be at the location of the maximum Sobel response.

This method has its advantages and disadvantages. GLOESS is effective at reducing false TRGB detections caused by noise and can help suppress smaller local maxima in the Sobel response to isolate the strongest discontinuities in a luminosity function. In addition, the method is computationally inexpensive compared to maximum likelihood estimation, for instance, and can be used to detect multiple TRGBs in a sample containing mixed stellar populations. However, the smoothing filter, by nature, reduces the TRGB discontinuity by moving red giant branch (RGB) stars fainter than the TRGB discontinuity into the region brighter than the TRGB discontinuity containing asymptotic giant branch (AGB) stars, as RGB stars are more abundant than AGB stars. This can cause a smoothing bias at certain levels of smoothing, where changing the strength of the smoothing changes the location of the measured TRGB. The amount of smoothing bias can also depend on the heights of the TRGB discontinuities. This effect was noted by Hatt et al. 2017 \cite{Hatt:2017rxl} and investigated and documented by Anderson et al. 2022 \cite{Anderson:2023aga} and Li et al. 2023 \cite{Li:2023utj}. Hatt et al. 2017 \cite{Hatt:2017rxl} describes an artificial star luminosity function test, which is used to choose a smoothing level by minimizing the quadrature sum of statistical and systematic errors. As the smoothing bias can be a function of the contrast ratio \cite{Anderson:2023aga, Li:2023utj}, it is essential that the artificial star luminosity function accurately reflects the observed sample for which is being used to estimate uncertainties. One should keep in mind that using different smoothing levels for different TRGB measurements in different galaxies could potentially lead to inconsistencies along the distance ladder. In addition, there can often be several local maximum in the Sobel response that are close together in height and magnitude even after applying GLOESS and Poisson weighting. Some studies suggest to increase the smoothing interval in these cases until there is a single, well defined peak (for instance, in Hatt et al. 2017 \cite{Hatt:2017rxl}). However, doing so runs the risk of combining Sobel peaks corresponding to multiple true TRGBs from different stellar populations or merging a local maximum corresponding to the true TRGB with another local maximum caused by noise. 

\subsubsection{Least Squares Fitting} \label{subsec:Least_Squares_Fitting}

The least-squares fitting method is frequently used by the Extragalactic Distance Database (EDD) team and has been used to measure $H_0$ \cite{Anand:2021sum}. This method assumes that the RGB/AGB luminosity function around the TRGB is well described by a broken power law model of the form:

\begin{equation}
\label{eq:LF}
\psi (m) = \begin{cases}
10^{a(m - m_{TRGB}) + b} & \quad m \ge m_{TRGB}\\
10^{c(m - m_{TRGB})}         & \quad m < m_{TRGB}.
\end{cases}
\end{equation}

\noindent where $m_{TRGB}$ is the magnitude of the TRGB, $m$ is the magnitude of a star, $a$ and $c$ are the logarithmic slopes of the luminosity function corresponding to the RGB and AGB, respectively, and $b$ represents the strength or height of the TRGB break. This form of the RGB/AGB luminosity function was first described by Zoccali \& Piotto 2000 \cite{Zoccali:2000ri} and subsequently used for TRGB measurements by Mendez et al. 2002, Makarov et al. 2006, Wu et al. 2014, Li et al. 2022, and Li et al. 2023, and references therein \cite{Mendez:2002ye, Makarov:2006wc, Wu:2014zxa, Li:2022aho, Li:2023pmo}. We note that Equation \ref{eq:LF} is also used for the maximum likelihood approach to measuring the TRGB, which is also used by the EDD team \cite{Makarov:2006wc} and has been used to measure the TRGB using Milky Way field giants \cite{Li:2022aho, Li:2023pmo}. Although both the least-squares and maximum likelihood methods are model-based, we focus here on the the least-squares fitting method here due to its use to measure $H_0$.

The TRGB least-squares fitting procedure as described in Wu et al. 2014 \cite{Wu:2014zxa} is as follows: The luminosity function model described in Equation \ref{eq:LF} is first convolved with the completeness, uncertainty, and bias of the observations. Then, an initial guess for the location of the TRGB is obtained by taking first derivatives of the luminosity function and locating their maximum. The luminosity function is binned in 0.05~mag intervals across $\pm$ 1~mag from the anticipated TRGB magnitude. Finally, a least-squares fit using the Levenberg-Marquardt algorithm is applied to the luminosity function to determine $m_{TRGB}$, $a$, $b$, and $c$. Uncertainties of the fitted parameters are estimated with the square root of the variance.

This method has the advantage of also being computationally inexpensive and is easily implemented in Python, with least-squares fitting routines such as from \texttt{scipy curve\_fit} \cite{Scipy:2019joe} readily available. However, this method assumes a single, true TRGB in its model and cannot separate out two TRGB breaks that may arise from the mixture of stellar populations. In principle, one could modify the model to include multiple TRGB discontinuities, however, this would still lack the flexibility of the Sobel filter at detecting multiple Sobel peaks. In addition, one must select an appropriate binning width; for well-populated stellar populations, the measured TRGB may be insensitive to binning, however, the measurements using sparsely populated populations may be susceptible to noise and sensitive to the choice of binning. The least-squares fit can also potentially converge onto unrealistic model parameters depending on the initial guess. For instance, in some cases the fit may identify the faint magnitude cutoff as the primary discontinuity if the TRGB discontinuity itself is not well defined or identify the tip of the AGB as the tip of the RGB (for examples, see Li et al. 2022 \cite{Li:2022aho} and Anand et al. 2019 \cite{Anand:2019}).

\begin{svgraybox}
    \textbf{FAQ}: Which method should I use to measure the TRGB?
    \newline \textbf{Response}: Each TRGB measurement method has its advantages and disadvantages and should be selected based on characteristic of the RGB/AGB sample and end goal. For instance, a Sobel filter paired with GLOESS can be more effective at reducing noise and false detections in noisy data compared to the binned power law least-squares fit, but the latter can also provide information about the RGB and AGB slopes, which may be useful for comparing stellar populations in other fields or galaxies. In many cases when the sample is well defined and populated, different methodologies agree. However, we emphasize that regardless of the method used, it is important to closely examine potential sources of measurement biases to ensure that the methods are consistently across a distance ladder such that any biases do not propagate into the final distance or $H_0$ measurement.
\end{svgraybox}

%
\begin{figure}[b]
\sidecaption
\centering
\includegraphics[scale=.5]{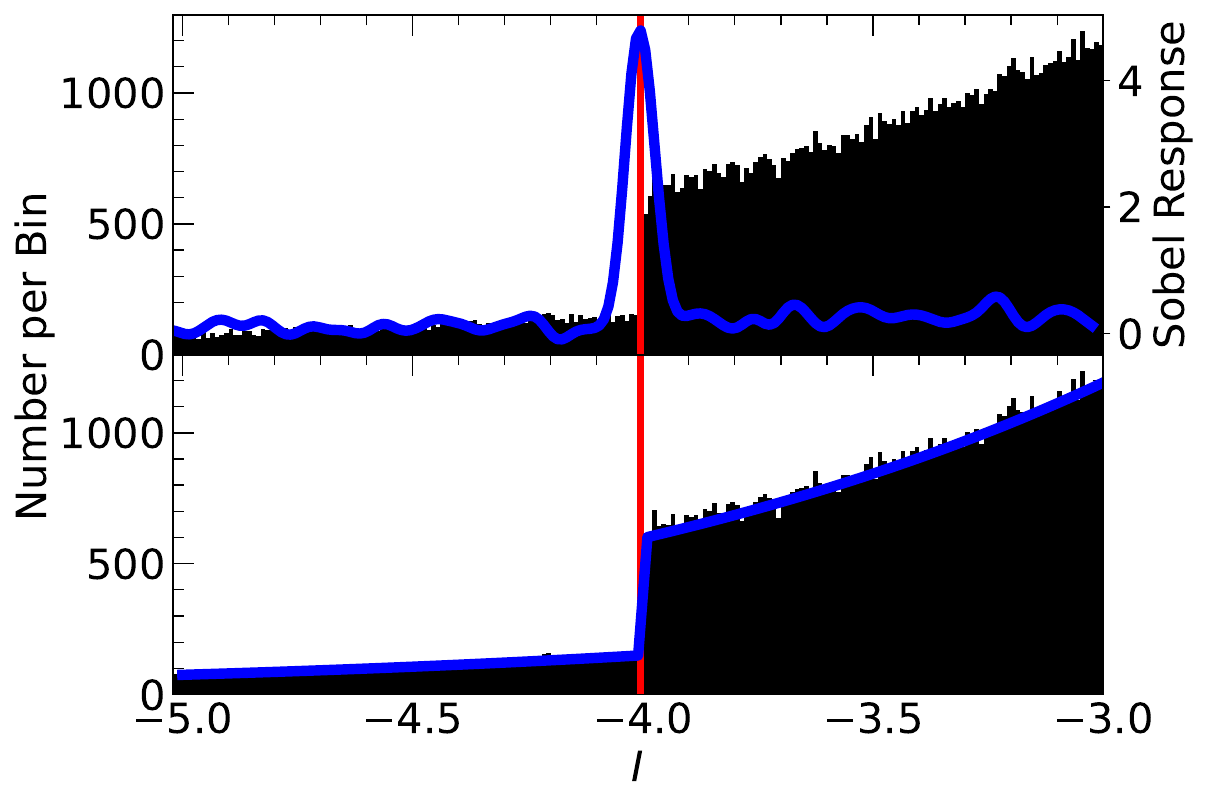}
%
%
\caption{Examples of a Sobel filter response (top; blue) and broken power law fit (bottom; blue) TRGB measurements on a simulated luminosity function (black). The red vertical line marks the location of the simulated TRGB.}
\label{fig:TRGB_Measurement_Example}       
\end{figure}

\subsubsection{Combining Algorithms and Stellar Populations} \label{subsec:Combining Algorithms and Stellar Populations}


The TRGB measurement methods described in Section \ref{subsec:Different Measurement Algorithms} have been used extensively to measure extragalactic distances and $H_0$. However, significant variations in calibrated TRGB luminosities (see compilations in Blakeslee et al. 2021, Freedman 2021, and Li et al. 2022 \cite{Blakeslee:2021rqi, Freedman:2021ahq, Li:2022aho}) and $H_0$ (for instance, see Freedman 2021 and Anand et al. 2021 \cite{Freedman:2021ahq, Anand:2021sum}) suggest there may be additional considerations that need to be taken into account to improve the consistency of TRGB measurements. In particular, one must ensure that the TRGB calibrated in an anchor galaxy is standarized such that it accurately reflects the true luminosity of the TRGB in a host galaxy. Mismatches between the TRGBs in anchor and host galaxies can occur with inconsistent measurement or selection choices in the two galaxies. The stellar population used to calibrate the TRGB in a host galaxy should have similar characteristics or be corrected to a fiducial value. One must also be careful that the measurement method also does not introduce biases, or that these biases cancel out along the distance ladder.
    
Some of these issues can manifest when using Sobel filter and least-squares fit TRGB measurement methods. Model based methods, such as the least-squares fit described in Section \ref{subsec:Least_Squares_Fitting}, have the disadvantage of assuming only a single TRGB break. While such model based methods are effective at measuring populations that show a single, well defined TRGB discontinuity, a population consisting of stars of several characteristics may exhibit multiple TRGB discontinuities \cite{Wu:2022hxf, Scolnic:2023mrv, Li:2023utj}. Multiple TRGB discontinuities cannot be separated with a broken power law model that assumes a single TRGB discontinuity. On the other hand, it is difficult to differentiate between local maxima in a Sobel response that correspond to noise and true TRGBs. An approach adopted by Hatt et al. 2017 \cite{Hatt:2017rxl} is to increase the smoothing until a single local maximum in the Sobel response remains. However, this can potentially combine multiple true TRGBs or a true TRGB with nearby local maxima in the Sobel response corresponding to noise, and different choices of smoothing in the calibration and host can lead to inconsistencies along the distance ladder. The selection of different smoothing parameters across the distance ladder, or one local maximum in the Sobel response over another of similar height, introduces subjective choices in the  measurement method that can easily lead to confirmation bias, if one is not careful.

To address these issues, the CATs team developed an unsupervised Sobel-filter based algorithm that can be used to measure TRGBs using objectively determined measurement parameters and improve the consistency of TRGB measurements across the distance ladder to measure $H_0$ \cite{Wu:2022hxf}. The algorithm first applies a step called spatial clipping, which removes regions in an observed field that contain high ratios of blue to red stars, as defined on a CMD, that might contaminate the red giant sample. Then, the algorithm applies a color band with a fixed width that maximizes the number of stars in the sample and applies GLOESS and Poisson weights to a luminosity function. A Sobel filter of the form [-1, 0, 1] is then evaluated across the luminosity function to identify the TRGB. The key differences in the CATs algorithm compared to previous Sobel filter algorithms are that the measurement parameters, such as the width of the color bands, choice of smoothing, and quality cuts, are first optimized to minimize the scatter in the observed field-to-field variation in the measured TRGBs in multiple fields from the Galaxy Halos, Outer disks, Substructure, Thick disks, and Star clusters (GHOSTS) galaxies \cite{Radburn-Smith:2011}. These measurement parameters are subsequently fixed when measuring extragalactic distances and $H_0$. The algorithm also defines a contrast ratio, taken to be the ratio between number of stars in the 0.5~mag interval fainter and brighter than the measured TRGB break. In addition to being a useful for estimating ages \cite{Wu:2022hxf, Harmsen:2023}, the contrast ratio provides a metric for differentiating high and low signal TRGB discontinuities. A high contrast ratio corresponds to a well defined break, while a low contrast ratio corresponds to an ill-defined break that may be sensitive to noise. The CATs algorithm filters out contrast ratios less than 3 to avoid false detections \cite{Wu:2022hxf}. This algorithm is publicly available via the GitHub repository link provided in Section \ref{sec:Open_Datasets}. 
    
In addition to developing an unsupervised Sobel-based TRGB measurement algorithm, the CATs team discovered a 5 $\sigma$ relationship between the measured TRGB and contrast ratio (tip-contrast ratio, or TCR). This finding suggests that the TRGB can be further standardized via the contrast ratio so that it is used more consistently across the distance ladder. The CATs team used their unsupervised algorithm and TCR to calibrate the TRGB luminosity using NGC 4258 \cite{Li:2023utj} and measure $H_0$ \cite{Scolnic:2023mrv} and find a $H_0$ of 73.22 $\pm$ 2.06 km s$^{-1}$ Mpc$^{-1}$. There are four key differences in the CATs $H_0$ compared to previous Sobel-filter based $H_0$ measurements. First, the CATs algorithm uses measurement parameters such as smoothing and spatial clipping consistently across the distance ladder to facilitate first order cancellation of potential biases. Second, the CATs algorithm allows for the possibility of multiple true TRGB breaks in a luminosity function due to mixtures of stellar populations. Third, the CATs team corrects measured TRGBs to a fiducial contrast ratio to facilitate closer alignment between the anchor and host TRGB and to account for variations in the measured TRGB as seen in the TCR. Fourth, the CATs measurement follows a different treatment of SN Ia, such as applying peculiar velocity corrections (see Scolnic et al. 2023 \cite{Scolnic:2023mrv} for more details).

The CATs team hypothesize that the TCR is caused by a combination of astrophysical properties as well as the measurement process such as smoothing \cite{Wu:2022hxf, Scolnic:2023mrv, Li:2023utj}. However, the exact nature of the TCR is inconsequential to the validity of the trend; it is important to remember that the TCR is an \textit{empirical} relationship, such that, at the fundamental level, there is a variation in the measured TRGB that correlates with another parameter. The TRGB is corrected to account for this variation irrespective of the exact origin of the variation, whether due to astrophysical differences, measurement artifacts, or both. There are still improvements that can be made to this approach, and more studies can be conducted to better understand the nature of the TCR and whether the TCR exists when using other TRGB measurement methods. While there may be other avenues to further standardize the TRGB, these efforts show the beginning of a paradigm shift in TRGB measurements with the recognition that there are still small variations in measured TRGBs using the Sobel-filter method that were not previously accounted for, and that additional standardization should be applied to the measured TRGB as used to measure $H_0$.

\begin{svgraybox}
\textbf{FAQ}: How much does correcting for the contrast ratio change $H_0$?
\newline \textbf{Response: } Scolnic et al. 2023 \cite{Scolnic:2023mrv} compared the CATs $H_0$ result with previous $H_0$ measurements in their Table 5 and found that applying contrast ratio standardization increased $H_0$ from Freedman et al. 2021 \cite{Freedman:2021ahq} by 1.4 km s$^{-1}$ Mpc$^{-1}$ and decreased $H_0$ from Anand et al. 2022 \cite{Anand:2021sum} by 0.3 km s$^{-1}$ Mpc$^{-1}$ . It's worth noting that some of the largest differences between the different $H_0$ values also come from different treatments of SNe Ia. We refer the reader to Table 5 and its corresponding text in Scolnic et al. 2023  \cite{Scolnic:2023mrv} for more details.
\end{svgraybox}

\subsection{Open Datasets} \label{sec:Open_Datasets}


Open datasets and algorithms help improve transparency and community-wide discussions in the field. We list here currently public TRGB datasets and algorithms available through websites that we encourage the reader to explore in addition to datasets made directly available via journal publications. We encourage future studies to make their datasets and algorithms available to the public.

\begin{itemize}
    \item Photometry from the Extragalactic Distance Database (Tully et al. 2009, Anand et al. 2021 \cite{Tully:2009ir, Anand:2021}): \newline https://edd.ifa.hawaii.edu/
    
    \item Unsupervised Sobel-filter algorithm and results from the CATs team: (Wu et al. 2023, Scolnic et al. 2023, Li et al. 2023 \cite{Wu:2022hxf, Scolnic:2023mrv, Li:2023utj}): \newline https://github.com/JiaxiWu1018/Unsupervised-TRGB, https://github.com/JiaxiWu1018/CATS-$H_0$
    
    \item 2D Maximum Likelihood algorithm used to calibrate TRGB using field stars and $Gaia$: (Li et al. 2023 \cite{Li:2023pmo}): \newline https://github.com/siyangliastro/Gaia-DR3-Milky Way-TRGB
\end{itemize}

\subsection{TRGB at the Crossroads}



We pause here to highlight two distinct ideologies in constructing a TRGB distance ladder to measure $H_0$ that arise as a natural consequence of developments in observational abilities. The first we refer to as artisanal measurements, which involve adjusting measurement and selection parameters to optimize TRGB measurements. For instance, one may increase the smoothing parameter when using a Sobel filter method until there is a single and well defined TRGB detection, as suggested by Hatt et al. 2017 \cite{Hatt:2017rxl}, with the optimal smoothing parameter being different for each galaxy. Advocates for this approach may argue that using different measurement and selection choices is necessary when using heterogeneous datasets having different levels of photometric uncertainties and sample size, for instance,  which requires different treatments to isolate the TRGB. We note that when doing so, it is especially important to understand and minimize potential biases of the measurement method when working with data of varying qualities.

On the other hand, applying different measurement and selection criteria runs the risk of introducing inconsistencies along the distance ladder. For instance, Anderson et al. 2023 and Li et al. 2023 \cite{Anderson:2023aga, Li:2023utj} demonstrate that different smoothing, as applied to the same luminosity function, introduces different biases. Advocates for this approach argue that measurements must be applied consistently across the distance ladder for all galaxies to ensure first order cancellation of any biases and minimize inconsistencies between calibrator and host measurements. This approach is more feasible when working with datasets with taken with the same telescope, such as $HST$ observations of NGC 4258 and GHOSTS galaxies.

These two ideologies can be seen in the context of a natural shift that occurs as more homogeneous observations are made possible with improved telescopes such as $HST$ and $JWST$. We leave the question of which approach is best open to the reader. Both have their advantages and disadvantages. While there is not a single, accepted answer, it is important to be aware of this distinction, especially when interpreting the results and methodologies of different $H_0$ results. 



\section{Future Developments} \label{sec:2}


\subsection{New Prospects with JWST} \label{sec:JWST}

\begin{figure}[b]
\sidecaption
\centering
\includegraphics[scale=.5]{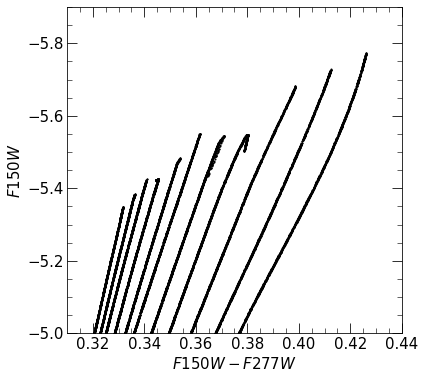}
%
%
\caption{Isochrones generated using the Python ArtPop package \cite{Greco:2022} in $JWST$ NIRCam $F150W$ and $F277W$ using stars of 10 Gyr and metallicities ranging from [Fe/H] of $-$2 to $-$1 dex.}
\label{fig:F150W_Isochrones}
\end{figure}

The successful launch of $JWST$ opened up the possibly of probing the TRGB further in the IR in greater resolutions and distances than $HST$. In addition, the configuration of detector modules on the $JWST$ Near Infrared Camera (NIRCam) opens up the possibility for observational programs to simultaneously observe multiple standard candles at once. NIRCam contains two cameras each consisting of four chips that creates a rectangular array. This setup allows $JWST$ to simultaneously image both the halos and disks of galaxies to observe red giants for TRGB, Cepheid variables, Mira variables, and the Carbon asymptotic giant branch stars for the J-region Asymptotic Giant Branch (JAGB), potentially enabling four $H_0$ measurements with one set of anchor and host observations once the JAGB is better understood and standardized. This setup also will allow for a powerful cross-calibration of these standard candles. In addition, NIRCam is a two-channel instrument which allows observations to be taken simultaneously in the short and long wavelength IR, further increasing the breadth of observations possible within a single observing program. However, as the TRGB exhibits a greater metallicity dependence in the IR compared to the $I$ or $F814W$ bands \cite{McQuinn:2019},  further calibration and investigation will be needed before the IR TRGB can be used to measure robust extragalactic distances. We provide an example of the metallicity dependence of the NIRCam $F150W$ TRGB in Fig. \ref{fig:F150W_Isochrones}, where we use the Python ArtPop package \cite{Greco:2022} to generate 10 Gyr isochrones for the red giant branch with metallicities ranging from [Fe/H] of $-$2 to $-$1 in 0.1 increments starting from the left most isochrone. $JWST$ programs designed with these features in mind have already been completed; we highlight some of these in Section \ref{sec:Early_JWST_Observational_Programs}. 

%

\subsubsection{Early $JWST$ Observational Programs} \label{sec:Early_JWST_Observational_Programs}


As of writing this chapter, $JWST$ has undergone two general observer (GO) proposal cycles in addition to the Director’s Discretionary (DD) Early Release Science Programs (ERS), all of which contain accepted proposals containing observations that can be used to measure the TRGB. We briefly highlight some of these programs that contain observations that can be used for TRGB measurements and encourage the reader to explore images from these proposals on MAST\footnote{\url{https://mast.stsci.edu/portal/Mashup/Clients/Mast/Portal.html}}. Proposal abstracts can be obtained from STScI\footnote{\url{https://www.stsci.edu/jwst/science-execution/approved-ers-programs}} \footnote{\url{https://www.stsci.edu/jwst/science-execution/approved-programs}}. We anticipate this list to grow as $JWST$ undergoes more proposal cycles in the future.

\begin{itemize}

    \item \textbf{ERS DD-1334 (PI: Daniel Weisz)}: The Resolved Stellar Populations Early Release Science Program
    
    \item \textbf{Cycle 1 GO-1638 (PI: Kristen McQuinn)}: Securing the TRGB Distance Indicator: A Pre-Requisite for a JWST Measurement of $H_0$
    
    \item \textbf{Cycle 1 GO-1685 (PIs: Adam Riess)}: Uncrowding the Cepheids for an Improved Determination of the Hubble Constant

    \item \textbf{Cycle 1 GO-1995 (PI: Wendy Freedman \& Barry Madore)}: Answering the Most Important Problem in Cosmology Today: Is the Tension in the Hubble Constant Real?

    \item \textbf{Cycle 2 GO-3055 (PI: Richard Brent Tully)}: A TRGB calibration of Surface Brightness Fluctuations

\end{itemize}

\subsection{Future Gaia Data Releases}


The $Gaia$ mission \cite{Gaia:2016zol} has observed over a billion stars in the Milky Way and significantly augmented the understanding of our home galaxy. Among the measurements of Milky Way field stars made by $Gaia$ include low-resolution spectra that can be used to construct synthetic photometry and parallaxes. These measurements can be retrieved via the $Gaia$ archive\footnote{\url{https://gea.esac.esa.int/archive/}} and $Gaia$ Synthetic Photometry Catalogue \cite{Gaia:2023} and have been used to calibrate the TRGB. Dixon et al. 2023 \cite{Dixon:2023jeo} used $Gaia$ Early Data Release (EDR) 3 parallaxes and SkyMapper Data Release (DR) 3, American Association of Variable Star Observers (AAVSO) Photometric All-Sky Survey (APASS) DR9, Asteroid Terrestrial-impact Last Alert System (ATLAS) Refcat2, and $Gaia$ DR3 synthetic photometry to calibrate the TRGB using Milky Way field stars. They directly convert apparent magnitudes into absolute magnitudes using distances constructed from $Gaia$ parallaxes and apply a Sobel filter to find a weighted average calibration of $M^{TRGB}_I$ = $-$4.042 $\pm$ 0.041 (stat) $\pm$ 0.031 (sys). Caution should be used, however, when directly converting apparent magnitudes to absolute magnitudes via parallaxes due to asymmetric probability distributions (see Bailer-Jones et al. 2021 \cite{Bailer-Jones:2021}). Li et al. developed a two-dimensional maximum likelihood estimation algorithm to ensure simultaneous treatment of uncertainties in parallaxes and magnitudes. The algorithm was first introduced in Li et al. 2022 \cite{Li:2022aho} and optimizes model parameters for a broken power law luminosity function and exponential density distribution to find the set of parameters that maximizes the likelihood of observing the sample of field giants. This first study used $Gaia$ EDR3 parallaxes and APASS DR9 photometry and was subsequently improved using $Gaia$ DR3 synthetic photometry in Li et al. 2023 \cite{Li:2023pmo} to find a calibration of $M^{TRGB}_I$ = $-$3.970 $^{+0.042}_{-0.024}$ (sys) $\pm$ 0.062 (stat). The algorithm used for these two studies by Li et al. is publicly available via the GitHub repository link provided in Section \ref{sec:Open_Datasets}. The differences between the Dixon et al. 2023 \cite{Dixon:2023jeo} and Li et al. 2023 \cite{Li:2023pmo} field star results can be due to several factors, such as a different treatment of the $Gaia$ parallax zero-point offset (differences of 10 $\mu$as), which is still not well understood, and differences in measurement methodologies, with one being in magnitude space and the other in parallax space.  

$Gaia$ parallaxes can also be used to calibrate the TRGB in globular clusters such as $\omega$ Centauri. Soltis et al. 2021 \cite{Soltis:2020gpl} used $Gaia$ EDR3 parallaxes to calibrate the TRGB measured by Bellazzini et al. 2001 \cite{Bellazzini:2001fg} and find $M_{I,TRGB}$ = $-$3.97 $\pm$ 0.06~mag. Li et al. 2023 \cite{Li:2023pmo} investigated the possibility of using $Gaia$ DR3 synthetic photometry to measure the TRGB in $\omega$ Centauri but found evidence of blending that prevents it from being used to obtain an accurate TRGB calibration. They instead update the TRGB measurement in $\omega$ Centauri using the database from Stetson et al. 2019 \cite{Stetson:2019} to find $M_{I,TRGB}$ = $-$3.97 $\pm$ 0.04 (stat) $\pm$ 0.10 (sys) mag.

Higher precision parallaxes and synthetic photometry in future $Gaia$ data releases can further improve these TRGB calibrations. We note that a better understanding of the $Gaia$ parallax zero-point offset will be essential before $Gaia$ data can be used to obtain a TRGB calibration that is competitive in precision with calibrations using other anchors such as NGC 4258. At the moment, the parallax zero-point offset is not well understood; the standard offset is described in Lindegren et al. 2021 \cite{Lindegren:2021}, but other estimates can differ \cite{Groenewegen:2021, Huang:2021, Ren:2021, Zinn:2021, Maiz_Apellaniz:2022}. Regardless, $Gaia$ has the potential to help increase leverage in TRGB $H_0$ calibrations through observations independent from those used for other anchors.

\begin{acknowledgement}
The authors would like to thank Dr. Gagandeep Anand for helpful comments. S.L. is supported by the National Science Foundation Graduate Research Fellowship Program under grant number DGE2139757. 
R.L.B. is supported by the National Science Foundation through grant number AST-2108616.
\end{acknowledgement}

\end{document}